\documentstyle[A4,epsfig]{article}

\title{Quantum dynamics and statistics of two coupled down-conversion
processes}
\author{Ladislav Mi\v{s}ta, Jr.
\thanks{e-mail mista@optnw.upol.cz},
 Ji\v{r}\'{\i} Herec
\thanks{e-mail herec@optnw.upol.cz},
 Viktor Jel\'{\i}nek
\thanks{e-mail jelinek@risc.upol.cz},\\
 Jaroslav \v{R}eh\'{a}\v{c}ek
\thanks{e-mail rehacek@alpha.inf.upol.cz},
Jan Pe\v{r}ina
\thanks{Also Joint Laboratory of Optics, Palack\'{y}
University and Physical Institute of Academy of Sciences of the Czech
Republic, Olomouc, Czech Republic.}
\thanks{e-mail perina@risc.upol.cz}
\\
Department of Optics,
Palack\'{y} University,\\
17. listopadu 50, 772 00 Olomouc,\\
Czech Republic}
\date{}

\begin{document}
\maketitle

\begin{abstract}

In the framework of Heisenberg-Langevin theory the dynamical and statistical
effects arising from the linear interaction of two nondegenerate down-conversion
processes are investigated. Using the strong-pumping approximation the
analytical solution of equations of motion is calculated. The phenomena
reminiscent of Zeno and anti-Zeno effects are examined. The possibility of
phase-controlled and mismatch-controlled switching is illustrated.

\end{abstract}

\section{Introduction}
\qquad Optical parametric processes yield a wide variety of optical
phenomena. It is not surprising that many new phenomena will arise
if a parametric process is coupled to another one or to a different
optical process. For instance, the superposition of signal photons originating
from two down-convertors with aligned idler beams leads to nontrivial
quantum interference effects \cite{Mandel}. Parametric process coupled
via Kerr interaction to an auxiliary mode, exhibiting quantum Zeno effect is
another nice example \cite{Luis,Soto}. Many such composite systems (usually
called nonlinear couplers) has thoroughly been studied in the literature.
All-optical switching in the assymetric nonlinear coupler operating by the
second-harmonic generation has been investigated in \cite{Asant} and its
non-classical behaviour has been discussed in \cite{Per1,Per2}.
The quantum dynamics and statistics of the symmetric coupler containing
two second-harmonic processes have been examined in \cite{Per3}. The
coupler composed of one linear waveguide and one nonlinear waveguide
operating by the down-coversion process has been investigated in
\cite{Janszky} from the point of view of all optical switching. The
occurrence of quantum Zeno and anti-Zeno effects in a similar device has
been reported in \cite{Reh}. Amplitude behaviour of two linearly coupled
down-conversion processes has been studied in \cite{Janszky}. Short-length
analysis of this device has been given in \cite{Herec}.

In this paper we deal with interesting phenomena
arising as a consequence of linear interaction between beams
propagating through the symmetric nonlinear coupler, which is composed of
two nonlinear waveguides based on the down-conversion processes. In fact
this arrangement can be looked at as a continuous version of famous Mandel's
experiment \cite{Mandel}, involving real physical interaction between the two
down-conversion processes. In Section~\ref{sec_sol} the equations of motion
are derived and their analytical solutions are given. Sections~\ref{sec_dyn}
and \ref{sec_stat} are devoted to the study of quantum dynamics and statistics
of the coupler. Its non-classical properties are discussed in
Section~\ref{sec_disc}.
%%%%%%%%%%%%%%%%%%%%%%%%%figure1%%%%%%%%%%%%%%%%%%%%%%%%%%%%%
\begin{figure}
\vspace{-2.7cm}
\centerline{\hspace{0cm}\psfig{width=17cm,angle=0,file=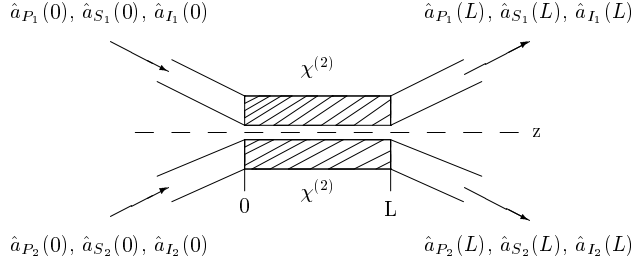}}
\vspace{-16cm}
\caption{Sketch of symmetric nonlinear coupler formed from two nonlinear
waveguides with susceptibility $\chi^{(2)}$. The interacting beams are
described by annihilation operators; $L$ is the interaction length.}
\label{fig_sket}
\end{figure}
%%%%%%%%%%%%%%%%%%%%%%%%%%%%%%%%%%%%%%%%%%%%%%%%%%%%%%%%%%%%%
\section{Equations of motion and their solution}\label{sec_sol}

\qquad  The coupler which is investigated in this article is composed of
two nonlinear waveguides operating by the down-conversion processes in a
directional arrangement (see Fig.~\ref{fig_sket}). The linear energy exchange
by means of evanescent waves between pump, signal and idler beams is considered.
The nonlinear media are assumed to be lossy. If such a system is far from
resonance, the effective description involving only the field variables
is adequate \cite{Graham1}. The effective momentum operator then reads
%%%%%%%%%%%%%%%%%%%%%%%%%%%%%%%%%%%%%%%%%%%%%%%%%%%%%%%%%%%%%
\begin{equation}\label{mom}
\hat{G}=\hat{G}_1+\hat{G}_2+\hat{G}_{res.}+\hat{G}_{res.-syst.}+
\hat{G}_{int.},
\end{equation}\\
where
\begin{eqnarray}\label{partmoment}
\hat{G}_i&=&\hbar\sum_{j={P_i},{S_i},{I_i}}k_j\hat{a}_{j}^{\dag}
\hat{a}_j+\hbar\left(\Gamma_i\hat{a}_{P_i}\hat{a}_{S_i}^{\dag}
\hat{a}_{I_i}^{\dag}+\mbox{h.c.}\right)\mbox{\qquad{for\quad{i=1,2}}},
\nonumber\\
\hat{G}_{res.}&=&\hbar\sum_{i=1}^{2}\sum_{j={P_i},{S_i},{I_i}}\sum_{l}
k_{lj}\hat{b}_{lj}^{\dag}\hat{b}_{lj},\nonumber\\
\hat{G}_{res.-syst.}&=&-\hbar\sum_{i=1}^{2}\sum_{j={P_i},{S_i},{I_i}}
\sum_{l}\left(\kappa_{lj}\hat{a}_j\hat{b}_{lj}^{\dag}+\mbox{h.c.}\right),
\nonumber\\
\hat{G}_{int.}&=&\hbar(\kappa_{P}\hat{a}_{P_1}\hat{a}_{P_2}^{\dag}
+\kappa_{S}\hat{a}_{S_1}\hat{a}_{S_2}^{\dag}+\kappa_{I}\hat{a}_{I_1}
\hat{a}_{I_2}^{\dag}+\mbox{h.c.}),\
\end{eqnarray}
%%%%%%%%%%%%%%%%%%%%%%%%%%%%%%%%%%%%%%%%%%%%%%%%
where $\hat{a}_j({\hat{a}_j}^{\dag})$, $j=P_1,P_2,S_1,S_2,I_1,I_2$
are annihilation (creation) operators of pump, signal, and idler modes.
Corresponding wavevectors along the z-axis of propagation are $k_{P_1}$,
$k_{P_2}$, $k_{S_1}$, $k_{S_2}$, $k_{I_1}$ and $k_{I_2}$. Linear
coupling constants between pump, signal and idler modes are denoted
$\kappa_P$, $\kappa_S$ and $\kappa_I$. Nonlinear coupling constants
are denoted as $\Gamma_1$ and $\Gamma_2$. Each mode $j$ is coupled via
linear coupling constant $\kappa_{lj}$ to the $l$-th reservoir mode
characterized by annihilation (creation) operators $\hat{b}_{lj}
({\hat{b}_{lj}}^{\dag})$ and wavevector $k_{lj}$ along z-axis of propagation.
The symbol $\hbar$ denotes the reduced Planck constant and h.c. represents
Hermitian conjugate terms.

The model represented by the momentum operator (\ref{mom}) is
symmetric  both under the exchange $1\leftrightarrow2$,
$\kappa_P\leftrightarrow\kappa_P^\ast$, $\kappa_S\leftrightarrow
\kappa_S^\ast$, $\kappa_I\leftrightarrow\kappa_I^\ast$ and under the
exchange $S\leftrightarrow I$. Since the dynamical behaviour of
the system is completely determined by its momentum operator, the
symmetries are conserved during evolution. This is convenient because it
is not necessary to write down all calculated quantities, the rest being
simply obtained by the above mentioned exchanges.

Substituting (\ref{mom}) into the Heisenberg equations of motion
$(i\hbar\frac{d}{dz}\hat{a}=[\hat{G},\hat{a}])$, introducing slowly
varying operators $\hat{A}_j(z)=\hat{a}_j(z)\mbox{exp}(-ik_jz)$ and
applying the Wigner-Weisskopf approximation \cite{Meystre}, we arrive
at the following Heisenberg-Langevin equations of motion
%%%%%%%%%%%%%%%%%%%%%%%%%%%%%%%%%%%%%%%%%%%%%%%%%
\begin{eqnarray}\label{acka}
\frac{d\hat{A}_{P_1}}{dz}&=&-\gamma_{P_1}\hat{A}_{P_1}+i{\kappa}_{P}^{\ast}
\hat{A}_{P_2}\mbox{exp}(-i\Delta k_Pz)+i{\Gamma}_{1}^{\ast}\hat{A}_{S_1}
\hat{A}_{I_1}\mbox{exp}(-i\Delta l_1z)+\hat{L}_{P_1}(z),\nonumber\\
\frac{d\hat{A}_{S_1}}{dz}&=&-\gamma_{S_1}\hat{A}_{S_1}+i{\kappa}_{S}^{\ast}
\hat{A}_{S_2}\mbox{exp}(-i\Delta k_Sz)+i\Gamma_1\hat{A}_{P_1}\hat{A}_{I_1}
^\dag\mbox{exp}(i\Delta l_1z)+\hat{L}_{S_1}(z),\nonumber\\
\end{eqnarray}
%%%%%%%%%%%%%%%%%%%%%%%%%%%%%%%%%%%%%%%%%%%%%%%%
where  $\Delta k_k=k_{k_1}-k_{k_2}$, $k=P,S,I$ are linear mismatches,
$\Delta l_i=k_{P_i}-k_{S_i}-k_{I_i}$, $i=1,2$ are nonlinear mismatches,
$\gamma_j$, $j=P_1,P_2,S_1,S_2,I_1,I_2$ are damping contants and the Langevin
forces $\hat{L}_j(z)$ are assumed to be Markoffian
%%%%%%%%%%%%%%%%%%%%%%%%%%%%%%%%%%%%%%%%%%%%%%%%
\begin{eqnarray}\label{Langevin}
\langle\hat{L}_j(z)\rangle&=&\langle\hat{L}_{j}^\dag(z)\rangle=
\langle\hat{L}_j(z)\hat{L}_k(z')\rangle=0,
\nonumber\\
\langle\hat{L}_{j}^\dag(z)\hat{L}_k(z')\rangle&=&2\gamma_j\langle
n_{dj}\rangle\delta_{jk}\delta(z-z'),\nonumber\\
\langle\hat{L}_j(z)\hat{L}_{k}^\dag(z')\rangle&=&2\gamma_j(\langle
n_{dj}\rangle+1)\delta_{jk}\delta(z-z').
\end{eqnarray}
%%%%%%%%%%%%%%%%%%%%%%%%%%%%%%%%%%%%%%%%%%%%%%%%
Here angle brackets denote the averaging over the reservoirs, $\langle
n_{dj}\rangle$ is one-mode mean photon number of the $j$-th reservoir,
$\delta_{jk}$ is the Kronecker symbol and $\delta(z)$ is the Dirac delta
function.
%%%%%%%%%%%%%%%%%%%%%%%%%%%%%%%%%%%%%%%%%%%%%%%%
It is useful to introduce the auxiliary quantities
%%%%%%%%%%%%%%%%%%%%%%%%%%%%%%%%%%%%%%%%%%%%%%%%
\begin{equation}\label{pomoc}
K_{S_i}=\gamma_{S_i}-i\Delta K_{S_i},\quad K_{I_i}=\gamma_{I_i}+i\Delta
K_{I_i},\quad i=1,2,
\end{equation}
%%%%%%%%%%%%%%%%%%%%%%%%%%%%%%%%%%%%%%%%%%%%%%%%
where
%%%%%%%%%%%%%%%%%%%%%%%%%%%%%%%%%%%%%%%%%%%%%%%%
\begin{equation}\label{Deltk}
\Delta K_{S_{1,2}}=\frac{\Delta k\pm\Delta k_{S}}{2},\quad
\Delta K_{I_{1,2}}=\frac{\Delta k\pm\Delta k_{I}}{2}
\end{equation}
%%%%%%%%%%%%%%%%%%%%%%%%%%%%%%%%%%%%%%%%%%%%%%%%
and
%%%%%%%%%%%%%%%%%%%%%%%%%%%%%%%%%%%%%%%%%%%%%%%%
\begin{equation}\label{globalk}
\Delta k=\frac{1}{2}\sum_{i=1}^{2}(k_{S_i}+k_{I_i}-k_{P_i}).
\end{equation}
%%%%%%%%%%%%%%%%%%%%%%%%%%%%%%%%%%%%%%%%%%%%%%%%
The mismatch (\ref{globalk}) contains wavevectors of all modes and thus
characterizes the overall phase mismatch. This important quantity
will be called global mismatch in the following.

If we assume the pump modes $P_1$, $P_2$ are stimulated by the
classical strong coherent fields
%%%%%%%%%%%%%%%%%%%%%%%%%%%%%%%%%%%%%%%%%%%%%%%%
\begin{equation}\label{strong}
\hat{A}_{P_1}(z)\rightarrow\xi_{P_1}\mbox{exp}(-i\Delta k_Pz/2),
\quad \hat{A}_{P_2}(z)\rightarrow\xi_{P_2}\mbox{exp}(i\Delta k_Pz/2),
\end{equation}
%%%%%%%%%%%%%%%%%%%%%%%%%%%%%%%%%%%%%%%%%%%%%%%%%
the system of equations of motion, represented by (\ref{acka}), splits into
two independent sets. The first one corresponds to $\{\hat{A}_{S_1},
\hat{A}_{S_2}, \hat{A}_{I_1}^\dag, \hat{A}_{I_2}^\dag\}$ operators and the
second one corresponds to their adjoints. In what follows we will confine
ourselves to the first set. The special choice of the phases of classical
amplitudes (\ref{strong}) leads, after the substitutions
%%%%%%%%%%%%%%%%%%%%%%%%%%%%%%%%%%%%%%%%%%%%%%%
\begin{equation}\label{subst1}
\hat{A}_{S_{1}}(z)=\hat{C}_{S_{1}}(z)\mbox{exp}(-i\Delta K_{S_1} z),
\quad\hat{A}_{I_{1}}^\dag(z)=\hat{C}_{I_{1}}^\dag(z)\mbox{exp}
(i\Delta K_{I_1}z),
\end{equation}
%%%%%%%%%%%%%%%%%%%%%%%%%%%%%%%%%%%%%%%%%%%%%%
to the system of linear differential equations with constant
coefficients of the form
%%%%%%%%%%%%%%%%%%%%%%%%%%%%%%%%%%%%%%%%%%%%%%
\begin{eqnarray}\label{cecka}
\frac{d\hat{C}_{S_1}}{dz}&=&-K_{S_1}\hat{C}_{S_1}+i\kappa_S^\ast\hat{C}_{S_2}
+iG_1\hat{C}_{I_1}^\dag+\hat{\cal{L}}_{S_1}(z),\nonumber\\
\frac{d\hat{C}_{I_1}^\dag}{dz}&=&-K_{I_1}\hat{C}_{I_1}^\dag
-i\kappa_I\hat{C}_{I_2}^\dag-iG_1^\ast\hat{C}_{S_1}+\hat{\cal{L}}_{I_1}^\dag(z),
\end{eqnarray}
%%%%%%%%%%%%%%%%%%%%%%%%%%%%%%%%%%%%%%%%%%%%%%%
where
%%%%%%%%%%%%%%%%%%%%%%%%%%%%%%%%%%%%%%%%%%%%%%%
\begin{displaymath}\label{malel}
\hat{\cal{L}}_{S_1}(z)=\hat{L}_{S_1}(z)\mbox{exp}(i\Delta K_{S_1}z),
\quad\hat{\cal{L}}_{I_1}^\dag(z)=\hat{L}_{I_1}^\dag(z)\mbox{exp}
(-i\Delta K_{I_1}z)
\end{displaymath}
%%%%%%%%%%%%%%%%%%%%%%%%%%%%%%%%%%%%%%%%%%%%%%%
are modified Langevin forces, and $G_1=\Gamma_1\xi_{P_1}$,
$G_2=\Gamma_2\xi_{P_2}$ are rescaled nonlinear coupling constants.

The system of Eqs.~(\ref{cecka}) can be solved using the Laplace
transformation method and method of variation of constants. Returning to
the operators $\hat{A}_j$, the solution can be written in the following
matrix form
%%%%%%%%%%%%%%%%%%%%%%%%%%%%%%%%%%%%%%%%%%%%%%%
\begin{equation}\label{sol}
\hat{\bf A}(z)={\bf M}(z)[{\bf X}(z)\hat{\bf A}(0)+\hat{\bf R}(z)],
\end{equation}
%%%%%%%%%%%%%%%%%%%%%%%%%%%%%%%%%%%%%%%%%%%%%%%
where we have introduced the vector [$(\enspace)^T$ means the transposition]
%%%%%%%%%%%%%%%%%%%%%%%%%%%%%%%%%%%%%%%%%%%%%%%
\begin{equation}\label{env}
\hat{\bf A}(z)=(\hat{A}_{S_1}(z),\hat{A}_{S_2}(z),
\hat{A}_{I_1}^\dag(z),\hat{A}_{I_2}^\dag(z))^T,
\end{equation}
%%%%%%%%%%%%%%%%%%%%%%%%%%%%%%%%%%%%%%%%%%%%%%%
the vector of reservoir contribution
%%%%%%%%%%%%%%%%%%%%%%%%%%%%%%%%%%%%%%%%%%%%%%%
\begin{equation}\label{res}
\hat{\bf R}(z)=(\hat{R}_{S_1}(z),\hat{R}_{S_2}(z),
\hat{R}_{I_1}(z),\hat{R}_{I_2}(z))^T,
\end{equation}
%%%%%%%%%%%%%%%%%%%%%%%%%%%%%%%%%%%%%%%%%%%%%%%
the diagonal matrix of mismatches
%%%%%%%%%%%%%%%%%%%%%%%%%%%%%%%%%%%%%%%%%%%%%%%
\begin{equation}\label{mismatch}
{\bf M}(z)=\mbox{diag}(\mbox{exp}(-i\Delta K_{S_1}z),
\mbox{exp}(-i\Delta K_{S_2}z),\mbox{exp}(i\Delta K_{I_1}z),
\mbox{exp}(i\Delta K_{I_2}z))
\end{equation}
%%%%%%%%%%%%%%%%%%%%%%%%%%%%%%%%%%%%%%%%%%%%%%%
and the matrix of coefficients
%%%%%%%%%%%%%%%%%%%%%%%%%%%%%%%%%%%%%%%%%%%%%%%
\begin{equation}\label{xmatrix}
X_{ij}(z)=\sum_{k=1}^{4}(A_k)_{ij}\mbox{exp}(\lambda_k z)
\quad\mbox{for}\quad i,j=1,..\,,4,
\end{equation}
%%%%%%%%%%%%%%%%%%%%%%%%%%%%%%%%%%%%%%%%%%%%%%%
where
%%%%%%%%%%%%%%%%%%%%%%%%%%%%%%%%%%%%%%%%%%%%%%%
\begin{eqnarray}\label{amatrix}
{\bf A}_k=[\prod_{i\not=k=1}^{4}(\lambda_k-\lambda_i)]^{-1}
(\lambda_k^3{\bf a}+\lambda_k^2{\bf b}
+\lambda_k{\bf c}+{\bf d}),\quad k=1,..\,,4.
\end{eqnarray}
%%%%%%%%%%%%%%%%%%%%%%%%%%%%%%%%%%%%%%%%%%%%%%%
Four-dimensional matrices ${\bf b}$, ${\bf c}$, ${\bf d}$ can be found
in Appendix~\ref{appA} and ${\bf a}$ is unity matrix.
The quantities $\lambda_k$, $k=1,..\,,4$ in (\ref{xmatrix}) and
(\ref{amatrix}) are single roots of the polynomial
%%%%%%%%%%%%%%%%%%%%%%%%%%%%%%%%%%%%%%%%%%%%%%%
\begin{equation}\label{polynom}
\Delta=x^4+ax^3+bx^2+cx+d,
\end{equation}
%%%%%%%%%%%%%%%%%%%%%%%%%%%%%%%%%%%%%%%%%%%%%%%
with coefficients
%%%%%%%%%%%%%%%%%%%%%%%%%%%%%%%%%%%%%%%%%%%%%%%
\begin{eqnarray}\label{koeficienty}
a&=&\gamma_{S_1}+\gamma_{S_2}+\gamma_{I_1}+\gamma_{I_2},\nonumber\\
b&=&L_S+L_I+\bar L_1+\bar L_2+K_{S_1}K_{I_2}+K_{I_1}K_{S_2},\nonumber\\
c&=&L_SK_{I_2}+L_IK_{S_1}+{\bar L_1}K_{S_2}+{\bar L_2}K_{I_1}\nonumber\\
&&+|\kappa_S|^2K_{I_1}+|\kappa_I|^2K_{S_2}-|G_1|^2K_{I_2}
-|G_2|^2K_{S_1},\nonumber\\
d&=&K_{S_1}K_{I_1}K_{S_2}K_{I_2}+|\kappa_S|^2 K_{I_1}K_{I_2}
+|\kappa_I|^2 K_{S_1}K_{S_2}-|G_1|^2 K_{S_2}K_{I_2}\nonumber\\
&&-|G_2|^2 K_{S_1}K_{I_1}+|\kappa_S\kappa_I-G_1^\ast G_2|^2,
\end{eqnarray}
%%%%%%%%%%%%%%%%%%%%%%%%%%%%%%%%%%%%%%%%%%%%%%%
where
%%%%%%%%%%%%%%%%%%%%%%%%%%%%%%%%%%%%%%%%%%%%%%%
\begin{displaymath}\label{lka}
L_j=K_{j_1}K_{j_2}+|\kappa_j|^2,\quad j=S,I,\quad \bar L_i=K_{S_i}K_{I_i}
-|G_i|^2,\quad i=1,2.
\end{displaymath}
%%%%%%%%%%%%%%%%%%%%%%%%%%%%%%%%%%%%%%%%%%%%%%%
If the roots of polynomial (\ref{polynom}) are multiple, we can obtain
the solution using the same methods.
%%%%%%%%%%%%%%%%%%%%%%%%%%%%%%%%%%%%%%%%%%%%%%%
\section{Quantum dynamics}\label{sec_dyn}

To investigate the dynamical behaviour of the coupler, we have to find
roots of the characteristic polynomial (\ref{polynom}).

If the damping is neglected and perfect phase matching is assumed, the
polynomial is quadratic in $x^2$ and its roots are easy to find. In this
case we can obtain periodical solution, exponentially amplifying solution
or a combination of these two depending on the parameters of the process
\cite{Mista1}.

If either the losses are included and all phase mismatches are zero, or
losses are neglected and phase mismatches are retained, we arrive at the
fourth-order polynomial with real coefficients. The roots can be found
using the Cardan formulae; unfortunately they are of complicated form and
it is more convenient to solve for the roots numerically.

In the most general case we need to solve the fourth-order equation with
complex coefficients, a task, which can only be performed with the help
of a computer. However, there are certain physically realizable regimes
%%%%%%%%%%%%%%%%%%%%%%%%%%%%%%%%%%%%%%%%%%%%%%%
\begin{eqnarray}\label{condit}
\gamma_{S_1}&=&\gamma_{S_2}=\gamma_{S},\quad \gamma_{I_1}=\gamma_{I_2}=
\gamma_{I},\nonumber\\
\Delta k_{S}&=&\Delta k_{I}=0,\quad G_1=G_2\frac{\kappa_S^\ast|\kappa_I|}
{\kappa_I|\kappa_S|},
\end{eqnarray}
%%%%%%%%%%%%%%%%%%%%%%%%%%%%%%%%%%%%%%%%%%%%%%%
for which the general polynomial (\ref{polynom}) factorizes into two
second-order ones. Their roots are
%%%%%%%%%%%%%%%%%%%%%%%%%%%%%%%%%%%%%%%%%%%%%%%
\small
\begin{eqnarray}\label{roots}
\lambda_{1,2}&=&\frac{-[\gamma_S+\gamma_I+i(|\kappa_I|-|\kappa_S|)]
\pm\sqrt{[\gamma_S-\gamma_I+i(|\kappa_S|+|\kappa_I|+\Delta
k)]^2+4|G_1|^2}}{2},\nonumber\\
\lambda_{3,4}&=&\frac{-[\gamma_S+\gamma_I+i(|\kappa_S|-|\kappa_I|)]
\pm\sqrt{[\gamma_S-\gamma_I+i(|\kappa_S|+|\kappa_I|-\Delta
k)]^2+4|G_1|^2}}{2}.\nonumber\\
\end{eqnarray}
\normalsize
%%%%%%%%%%%%%%%%%%%%%%%%%%%%%%%%%%%%%%%%%%%%%%%
Neglecting the damping ($\gamma_S=\gamma_I=0$), we can examine the
influence of the global mismatch $\Delta k$ to the dynamics of the coupler:
%%%%%%%%%%%%%%%%%%%%%%%%%%%%%%%%%%%%%%%%%%%%%%%
\begin{enumerate}
  \item If $\Delta k \in (-2|G_1|,2|G_1|)$, then
 \begin{enumerate}
  \item for $|\kappa_S|+|\kappa_I| \in (0,|\Delta k|+2|G_1|)$
  all roots have the form $\lambda_j=a_j+ib_j$ with nonzero real and
  imaginary parts $a_j$ and $b_j$
  \item for $|\kappa_S|+|\kappa_I| \in (|\Delta k|+2|G_1|,+\infty)$
  all roots are purely imaginary.
 \end{enumerate}
  \item If $\Delta k \in (-\infty,-2|G_1|)\cup(2|G_1|,+\infty)$, then
 \begin{enumerate}
 \item for $|\kappa_S|+|\kappa_I| \in (|\Delta k|-2|G_1|,|\Delta k|+
  2|G_1|)$ all roots are the same as in 1(a)
  \item for $|\kappa_S|+|\kappa_I| \in (0,|\Delta k|-2|G_1|)\cup
  (|\Delta k|+2|G_1|,+\infty)$ all roots are the same as in 1(b).
 \end{enumerate}
\end{enumerate}
%%%%%%%%%%%%%%%%%%%%%%%%%%%%%%%%%%%%%%%%%%%%%%%
Assuming the symmetrical linear coupling, $|\kappa_S|=|\kappa_I|$, the
imaginary parts $b_j$ of $\lambda_j$ in cases 1(a) and 2(a) vanish
and all roots acquire real values. In what follows, if $a_j\not=0$
($a_j=0$), we will say that the coupler operates in the hyperbolic (elliptic)
regime.
%%%%%%%%%%%%%%%%figure2%%%%%%%%%%%%%%%%%%%%%%%%%
\begin{figure}
\vspace{-3cm}
\centerline{\hspace{0cm}\psfig{width=17cm,angle=0,file=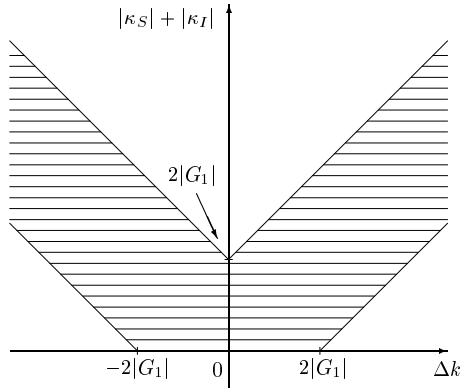}}
\vspace{-15cm}
\caption{Dependence of the character of dynamics of the coupler on
global mismatch $\Delta k$ and sum of linear coupling constants $|\kappa_S|
+|\kappa_I|$. Hatched area corresponds to the hyperbolic regime. The rest
corresponds to the elliptic regime.}
\label{fig_V}
\end{figure}
%%%%%%%%%%%%%%%%%%%%%%%%%%%%%%%%%%%%%%%%%%%%%%%%
A more instructive demonstration of regions of different dynamical
behaviour is given in Fig.~\ref{fig_V}.
%%%%%%%%%%%%%%%%%%%%%%%%%%%%%%%%%%%%%%%%%%%%%%%%
Notice first that $\Delta k$ and its counterpart $|\kappa_S|+|\kappa_I|$
affect the process in symmetrical ways. Now let us look closely at the
interplay between the linear coupling and global mismatch. If
$|\Delta k|<2|G_1|$ and the coupling strength is gradually increased,
one can see that the coupler crosses the border between the hyperbolic
regime (hatched area) and elliptic regime when a certain value of the linear
coupling strength $|\kappa_S|+|\kappa_I|$ is attained. More interestingly,
for large phase mismatch $|\Delta k|\gg2|G_1|$ (even such that almost no
energy is converted from the pump mode to the signal and idler modes), the
system moves up along the vertical line $\Delta k=\mbox{const.}$ in
Fig.~\ref{fig_V}, and it enters the region of instability characterized
by the domination of the down-conversion part of the evolution when
$|\kappa_S|+|\kappa_I|=\Delta k-2|G_1|$. For even stronger linear coupling
the coupler leaves the region of instability again (crossing the line
$|\kappa_S|+|\kappa_I|=\Delta k+2|G_1|$), the oscillatory character of the
evolution is restored and pump photons gradually cease to decay. Interpreting
(somewhat loosely) the linear coupling as a kind of continuous measuring
process, we can look at the just described behaviour as being a manifestation
of the well-known Zeno or anti-Zeno effects \cite{Luis,Soto}. Also here
the strong influence of the ``measuring apparatus'' leads to the hindering of
the decay of the originally unstable system. On the contrary under certain
conditions (here nonzero phase mismatch $\Delta k$), the decay of the unstable
system can be enhanced by a frequent (here continuous) monitoring of the
unstable system \cite{Soto,Reh}. Our Fig.~\ref{fig_V} clearly shows the
competition between these two opposite tendencies.
%%%%%%%%%%%%%%%%%%%%%%%%%%%%%%%%%%%%%%%%%%%%%%%%
\section{Quantum statistics}\label{sec_stat}

The quantum-statistical properties of the coupler are best studied
employing the normal characteristic function containing complete statistical
information about the system. The model represented by momentum operator
(\ref{mom}) together with the linearization procedure (\ref{strong}) lead
to the Gaussian characteristic function corresponding to the generalized
superposition of coherent fields and quantum noise \cite{Permon}
%%%%%%%%%%%%%%%%%%%%%%%%%%%%%%%%%%%%%%%%%%%%%%%%%
\begin{eqnarray}
C_{\cal N}(\{\beta_j\},z)&=&\mbox{exp}\left\{\sum_{j=1}^{4}
\left[-{B_j}(z)|\beta_j|^2+\frac{1}{2}({C_j}(z){\beta_j^\ast}^2+
\mbox{c.c.})+\right.\right.\nonumber\\
&&+\left.\left.\sum_{k=1,j<k}^{4}(D_{jk}(z){\beta}_j^
\ast{\beta}_k^\ast+\bar{D}_{jk}(z)\beta_j{\beta}_k^\ast+\mbox{c.c.})
\right.\right.\nonumber\\
&&+\left.\left.(\beta_j{\xi}_j^\ast(z)-\mbox{c.c.})\right]
\right\},
\end{eqnarray}
%%%%%%%%%%%%%%%%%%%%%%%%%%%%%%%%%%%%%%%%%%%%%%%%%
where the following identification $S_1\equiv 1$, $S_2\equiv 2$,
$I_1\equiv 3$, $I_2\equiv 4$ has been done. The complex amplitudes
$\xi_1(z)$, $\xi_2(z)$, $\xi_3(z)$, $\xi_4(z)$ are mean values of
operators $\hat{A}_{S_1}(z)$, $\hat{A}_{S_2}(z)$, $\hat{A}_{I_1}(z)$,
$\hat{A}_{I_2}(z)$, c.c. means the complex conjugated terms and
%%%%%%%%%%%%%%%%%%%%%%%%%%%%%%%%%%%%%%%%%%%%%%%%
\begin{eqnarray}\label{sumfce}
B_j(z)&=&\langle\Delta\hat{A}_{j}^\dag(z)\Delta\hat{A}_{j}
(z)\rangle,\quad C_j(z)=\langle(\Delta\hat{A}_{j}(z))^2\rangle,
\nonumber\\
D_{jk}(z)&=&\langle\Delta\hat{A}_{j}(z)\Delta\hat{A}_{k}
(z)\rangle,\quad \bar{D}_{jk}(z)=-\langle\Delta\hat{A}_{j}^\dag(z)
\Delta\hat{A}_{k}(z)\rangle
\end{eqnarray}
%%%%%%%%%%%%%%%%%%%%%%%%%%%%%%%%%%%%%%%%%%%%%%%%%
for $j,k=S_1,S_2,I_1,I_2$ are noise functions. The complicated explicit
expressions of the noise functions are given in Appendix~\ref{appB}.
The quantities $C_j=C_j(0)$ and $B_j=B_j(0)+1$ corresponding to the input
beams are expressed under the condition of independence of incident beams
in the form
%%%%%%%%%%%%%%%%%%%%%%%%%%%%%%%%%%%%%%%%%%%%%%%%%%%
\begin{equation}\label{noiseinit}
B_j=\mbox{cosh}^2(r_j)+\langle n_{chj}\rangle,\quad
C_j=\frac{1}{2}\mbox{exp}(i\theta_j)\mbox{sinh}(2r_j),
\end{equation}
%%%%%%%%%%%%%%%%%%%%%%%%%%%%%%%%%%%%%%%%%%%%%%%%%%%%
where $r_j$ and $\theta_j$, $j=1,2,3,4$ are squeeze parameters
and phases of the incident beams and $\langle n_{chj}\rangle$ represents
the mean number of external noise photons in the $j$-th mode.

Assuming the unsqueezed input fields ($r_j=0$) the explicit expressions
of noise functions (see Appendix~\ref{appB}) lead to the following identities
%%%%%%%%%%%%%%%%%%%%%%%%%%%%%%%%%%%%%%%%%%%%%%%%%%%%%
\begin{eqnarray}\label{identity}
C_{S_1}(z)&=&C_{S_2}(z)=C_{I_1}(z)=C_{I_2}(z)=0,\nonumber\\
D_{S_1S_2}(z)&=&\bar{D}_{S_1I_1}(z)=\bar{D}_{S_1I_2}(z)=0.
\end{eqnarray}
%%%%%%%%%%%%%%%%%%%%%%%%%%%%%%%%%%%%%%%%%%%%%%%%%%%%%
The quantum-statistical properties of single and compound modes can
be quantified by means of many statistical quantities. From these
we will use the principal squeeze variance $\lambda(z)$
[principal squeezing occurs if $\lambda<1(2)$ for single (compound)
mode], quadrature variances $\langle\left[\Delta{ \hat{q} \atop
\hat{p}}(z)\right]^2\rangle$ [quadrature squeezing occurs if $\langle
(\Delta \hat{q})^2 \rangle<1(2)$ or $\langle(\Delta \hat{p})^2\rangle<1(2)$
for single (compound) mode], normal reduced factorial moments of the
integrated intensity $\frac{\langle W^k(z)\rangle}{{\langle W(z)\rangle}^k}-1$
[they are negative for non-classical states, negative second moment reflects
the sub-Poissonian photon statistics] and the photon number distribution
$p(n,z)$ [quantum oscillations in $p(n,z)$ indicate the presence of state
having no classical analogy].

Adopting the standard definitions of the above mentioned quantities for
single mode \cite{Permon} and using (\ref{identity}), it is straightforward
to show that single modes do not exhibit any interesting behaviour.
In particular non-classical light cannot develop from coherent inputs in
single modes. It can arise only as a result of quantum correlations of
modes.

In the case of the compound mode $(i,j)$ the principal squeeze variance
$\lambda_{ij}(z)$, quadrature variances $\langle[\Delta \hat{q}_{ij}(z)]^2
\rangle$, $\langle[\Delta \hat{p}_{ij}(z)]^2\rangle$ and variance of the
integrated intensity $\langle[\Delta W_{ij}(z)]^2\rangle$ are defined as
follows
%%%%%%%%%%%%%%%%%%%%%%%%%%%%%%%%%%%%%%%%%%%%%%%%%%%%%%%%%%%%%%
\begin{equation}\label{lambda}
\lambda_{ij}(z)=2\{1+B_i(z)+B_j(z)-2\mbox{Re}[\bar D_{ij}(z)]-|C_i(z)+C_j(z)
+2D_{ij}(z)|\},
\end{equation}
%%%%%%%%%%%%%%%%%%%%%%%%%%%%%%%%%%%%%%%%%%%%%%%%%%%%%%%%%%%%%%
\begin{equation}\label{quadrature}
\langle(\Delta{ \hat{q} \atop \hat{p} })^2\rangle=2\{1+B_i(z)+B_j(z)
-2\mbox{Re}[\bar D_{ij}(z)]\pm\mbox{Re}[C_i(z)+C_j(z)+2D_{ij}(z)]\}
\end{equation}
%%%%%%%%%%%%%%%%%%%%%%%%%%%%%%%%%%%%%%%%%%%%%%%%%%%%%%%%%%%%%%
and
%%%%%%%%%%%%%%%%%%%%%%%%%%%%%%%%%%%%%%%%%%%%%%%%%%%%%%%%%%%%%%
\begin{equation}\label{intensity}
\langle[\Delta W_{ij}(z)]^2\rangle=\langle[\Delta W_i(z)]^2\rangle+
\langle[\Delta W_j(z)]^2\rangle+2\langle\Delta W_i(z)\Delta W_j(z)\rangle,
\end{equation}
%%%%%%%%%%%%%%%%%%%%%%%%%%%%%%%%%%%%%%%%%%%%%%%%%%%%%%%%%%%%%%
where
%%%%%%%%%%%%%%%%%%%%%%%%%%%%%%%%%%%%%%%%%%%%%%%%%%%%%%%%%%%%%%
\begin{eqnarray}\label{fluctint}
\langle[\Delta[W_i(z)]^2\rangle=B_i^2(z)+|C_i(z)|^2+2B_i(z)|\xi_i(z)|^2
+2\mbox{Re}[C_i(z){\xi_i^\ast}^2(z)]
\end{eqnarray}
%%%%%%%%%%%%%%%%%%%%%%%%%%%%%%%%%%%%%%%%%%%%%%%%%%%%%%%%%%%%%%%%
and
%%%%%%%%%%%%%%%%%%%%%%%%%%%%%%%%%%%%%%%%%%%%%%%%%%%%%%%%%%%%%%%%
\begin{eqnarray}\label{corelint}
\langle\Delta W_i(z)\Delta W_j(z)\rangle&=&2\mbox{Re}[D_{ij}(z)
\xi_i^\ast(z)\xi_j^\ast(z)-\bar D_{ij}(z)\xi_i(z)\xi_j^\ast(z)]
\nonumber\\
&&+|D_{ij}(z)|^2+|\bar D_{ij}(z)|^2.
\end{eqnarray}
%%%%%%%%%%%%%%%%%%%%%%%%%%%%%%%%%%%%%%%%%%%%%%%%%%%%%%%%%%%%%%
If the correlation function (\ref{corelint}) is negative, we say that
the corresponding modes are anti-correlated.

The definitions of sum photon number distribution and $k$-th moment
$\langle W_{ij}^k (z)\rangle$ are rather complex and can be found
in \cite{Perova}.
%%%%%%%%%%%%%%%%%%%%%%%%%%%%%%%%%%%%%%%%%%%%%%%%%%%%%%%%%%%%%%
\section{Discussion of results}\label{sec_disc}

As we have already mentioned above, analytical expressions of required
quantities are only available under certain simplifying and restrictive
assumptions. Even in those cases the expressions are of a complicated
form and thus almost useless for qualitative discussions. Therefore we will
employ numerical methods. The analytical solutions, when available,
may serve for checking the results of the numerical calculations.

This section is devoted to the investigation of interesting
phenomena arising from the linear coupling between two down-conversion
processes. Each phenomenon is discussed in separate subsection.
%%%%%%%%%%%%%%%%%%%%%%%%%%%%%%%%%%%%%%%%%%%%%%%%%%%%%%%%%%%%%%
\subsection{\it Quadrature switching}\label{sub_swit}

\begin{figure}
\vspace{-5.5cm}
\centerline{\hspace{0cm}\psfig{width=15cm,angle=0,file=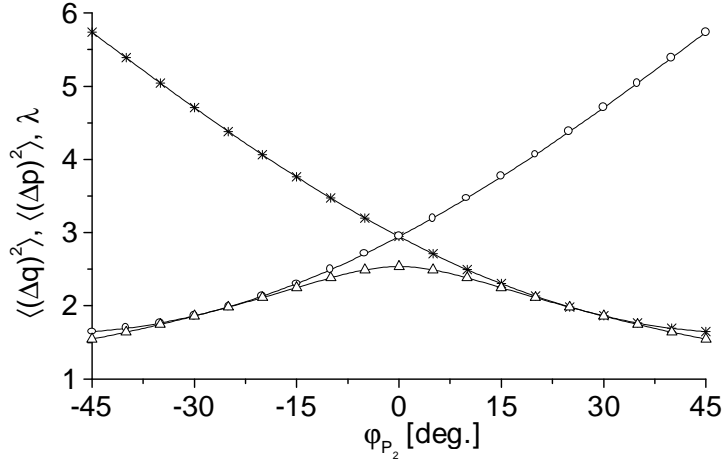}}
\vspace{-9.0cm}
\caption{Demonstration of phase-controlled switching between the quadrature
variances $\langle{\left[\Delta\hat{q}(L)\right]}^2 \rangle$ $(\circ)$ and
$\langle{\left[\Delta\hat{p}(L)\right]}^2 \rangle$ $(\ast)$ (the curve
denoted by $(\triangle)$ corresponds to the principal squeeze variance
$\lambda(L)$) of mode $(S_1,I_1)$; $L=1.2$, $\Gamma_1=\Gamma_2=1$,
$\kappa_S=\kappa_I=2$, $\Delta k=\Delta k_{S}=\Delta k_{I}=0$,
$\xi_{P_1}=|\xi_{P_2}|=1$, $\xi_{S_1}=\xi_{I_1}=\xi_{S_2}=\xi_{I_2}=0$,
$\gamma_j=0.2$, $\langle n_{dj} \rangle=10^{-2}$.}
\label{fig_swit}
\end{figure}

It was reported in \cite{Janszky} that the symmetric coupler $(|G_1|=|G_2|)$
where only signal modes are linearly coupled $(\kappa_I=0)$ behaves as follows.
If mode $S_1$ is squeezed in the given quadrature at the input, squeezing in a
conjugated quadrature develops in mode $S_2$. Taking into account also linear
exchange between idler modes, we can observe a similar phenomenon in quadratures
of compound mode $(S_1,I_1)$. Let us assume that both down-conversion processes
are spontaneous, linear coupling constants are symmetric $\kappa_S=\kappa_I$ and
sufficiently strong. Changing now the phase $\varphi_{P_2}\equiv\mbox{arg}\,
\xi_{P_2}$ of the pump mode $P_2$ and leaving the phase $\varphi_{P_1}\equiv
\mbox{arg}\,\xi_{P_1}$ of the pump mode $P_1$ fixed, we can switch between
quadratures at the output of mode $(S_1,I_1)$. Moreover, if the interaction
length $L$ is appropriatelly chosen, squeezing in the given quadrature can be
transferred to the conjugated one in a continuous way
(see Fig.~\ref{fig_swit}).
%%%%%%%%%%%%%%%%%%%%%%%%%figure3%%%%%%%%%%%%%%%%%%%%%%%%%%%%%%
%%%%%%%%%%%%%%%%%%%%%%%%%%%%%%%%%%%%%%%%%%%%%%%%%%%%%%%%%%%%%%%%%%%
\subsection{\it Linear coupling can compensate wrong phases}

\begin{figure}
\vspace{-5.3cm}
\centerline{\hspace{0cm}\psfig{width=17cm,angle=0,file=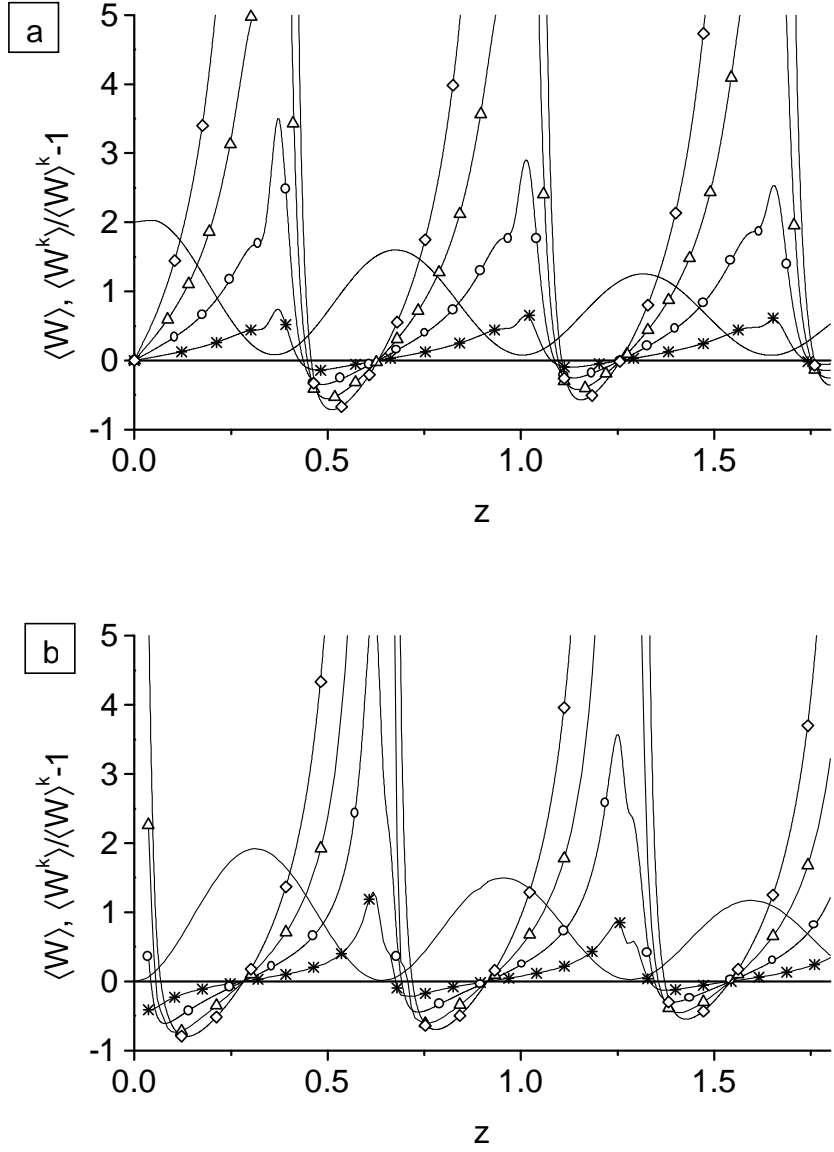}}
\vspace{-5cm}
\caption{The integrated intensity $\langle W(z)\rangle$ (---) and its
reduced factorial moments $\frac{\langle W^k(z)\rangle}
{{\langle W(z)\rangle}^k}-1$ for $k=2$ $(\ast)$, $k=3$ $(\circ)$,
$k=4$ $(\triangle)$, $k=5$ $(\diamond)$ for mode $(S_1,I_1)$ $(a)$ and
$(S_2,I_2)$ $(b)$; $\Gamma_1=\Gamma_2=1$, $\kappa_S=\kappa_I=5$,
$\Delta k=\Delta k_{S}=\Delta k_{I}=0$, $\xi_{P_1}=\xi_{P_2}=-i$,
$\xi_{S_1}=\xi_{I_1}=1$, $\xi_{S_2}=\xi_{I_2}=0$, $\gamma_j=0.2$,
$\langle n_{dj}\rangle=10^{-2}$.}
\label{fig45}
\end{figure}

\begin{figure}
\vspace{1cm}
\centerline{\hspace{0cm}\psfig{width=8cm,angle=0,file=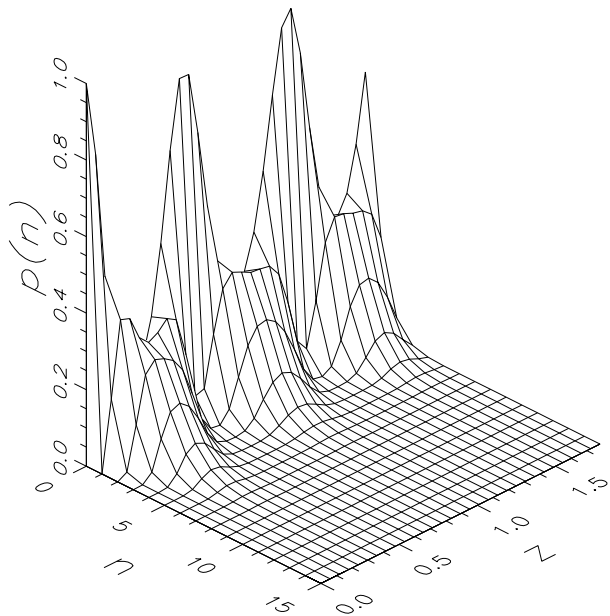}}
\vspace{0cm}
\caption{The sum photon number distribution $p(n,z)$ for mode $(S_2,I_2)$;
the parameters of the process are the same as in Fig.~\ref{fig45}.}
\label{fig_pn}
\end{figure}

It is well known \cite{Permon} that for small interaction lengths $z$
sub-Poissonian light can be generated in a nondegenerate
down-conversion process in compound mode $(S,I)$ provided that the process
is stimulated $(\xi_S,\xi_I\neq 0)$ and phases of incident beams fulfil
the optimum phase condition $\mbox{arg}(\xi_S\xi_I\xi_P^\ast)=-\frac{\pi}{2}$.
On the other hand, if either the process is spontaneous or the phase
condition is strongly violated, this mode is super-Poissonian. Let us
assume that the process in the first waveguide is stimulated by amplitudes
$\xi_{S_1},\xi_{I_1}$ strongly violating the optimum phase condition
(say $\mbox{arg}(\xi_{S_1}\xi_{I_1}\xi_{P_1}^\ast)=\frac{\pi}{2}$) and the
process in the second waveguide is spontaneous $(\xi_{S_2}=\xi_{I_2}=0)$.
Introducing the linear coupling between the waveguides, the modes
$(S_1,I_1)$ and $(S_2,I_2)$ can exhibit an interesting non-classical behaviour.
The linear coupling restores the optimum phase condition and sub-Poissonian
light is generated in mode $(S_1,I_1)$, surprisingly, for larger $z$ (see
Fig.~\ref{fig45}~(a)). Further, sub-Poissonian light is also generated in
mode $(S_2,I_2)$ for small $z$ (Figures~\ref{fig45}~(b) and \ref{fig_pn}),
a phenomenon, which cannot be explained as easy as in the previous case.
To deepen insight into this phenomenon we will resort to the analytical
results. If losses are neglected, all mismatches are zero,
$\kappa\equiv\kappa_S=\kappa_I$ is real, $G\equiv G_1=G_2$, and $\kappa>|G|$,
we can calculate, using the results of Section~\ref{sec_sol} and
(\ref{corelint}), the correlation of fluctuations
%%%%%%%%%%%%%%%%%%%%%%%%%%%%figures4(a),(b)%%%%%%%%%%%%%%%%%%%%%%%%
%%%%%%%%%%%%%%%%%%%%%%%%%%%%%%%%%%%%%%%%%%%%%%%%%%%%%%%%%%%%%%%%%%%%
\begin{eqnarray}\label{corelS2I2}
\langle\Delta W_{S_2}(z)\Delta W_{I_2}(z)\rangle&=&|G|^2u^2(z)v^2(z)
+2\kappa^2|G||\xi_{S_1}||\xi_{I_1}|u(z)v^3(z)\nonumber\\
&&\times\mbox{sin}(\varphi_P-\varphi_{S_1}-\varphi_{I_1}),
\end{eqnarray}
%%%%%%%%%%%%%%%%%%%%%%%%%%%%%%%%%%%%%%%%%%%%%%%%%%%%%%%%%%%%%%
where $\varphi_P$, $\varphi_{S_1}$, $\varphi_{I_1}$ are the phases of input
coherent amplitudes $\xi_P\equiv\xi_{P_1}=\xi_{P_2}$, $\xi_{S_1}$,
$\xi_{I_1}$ and
%%%%%%%%%%%%%%%%%%%%%%%%%%%%%%%%%%%%%%%%%%%%%%%%%%%%%%%%%%%%%%
\begin{eqnarray}\label{uvfce}
u(z)&=&\mbox{cos}\left[\sqrt{2\left(\kappa^2-|G|^2\right)}z\right]
-\frac{z}{4}\sqrt{2\left(\kappa^2-|G|^2\right)}\,\mbox{sin}
\left[\sqrt{2\left(\kappa^2-|G|^2\right)}z\right],\nonumber\\
v(z)&=&\frac{3\,\mbox{sin}\left[\sqrt{2\left(\kappa^2-|G|^2\right)}z
\right]}{4\left[\sqrt{2\left(\kappa^2-|G|^2\right)}\right]}+
\frac{z}{4}\,\mbox{cos}\left[\sqrt{2\left(\kappa^2-|G|^2\right)}
z\right].
\end{eqnarray}
%%%%%%%%%%%%%%%%%%%%%%%%%%%%%%%%%%%%%%%%%%%%%%%%%%%%%%%%%%%%%%%%%%%%
Since both the expressions on the right hand side (R.H.S.) of
Eqs.~(\ref{uvfce}) are real, only the second term on the R.H.S. of
Eq.~(\ref{corelS2I2}) can be negative depending on the sign of the product
$u(z)v(z)$, and on the argument of sine function. Restricting ourselves to
small $z$, we can expand $u(z)$ and $v(z)$ up to the $z^3$ around the origin,
and approximate Eq.~(\ref{corelS2I2}) by the expression
%%%%%%%%%%%%%%%%%%%%%%%%%%%%%%%%%%%%%%%%%%%%%%%%%%%%%%%%%%%%%%%%%%%%
\begin{equation}\label{shortcorelS2I2}
\langle\Delta W_{S_2}(z)\Delta W_{I_2}(z)\rangle\approx|G|^2z^2
+2\kappa^2|G||\xi_{S_1}||\xi_{I_1}|z^3\mbox{sin}(\varphi_P-
\varphi_{S_1}-\varphi_{I_1}).
\end{equation}
%%%%%%%%%%%%%%%%%%%%%%%%%%%%%%%%%%%%%%%%%%%%%%%%%%%%%%%%%%%%%%%%%%%%%
Note first, that anti-correlation can only arise for sufficiently strong
$\kappa$ and for sufficiently large $z$. It attains its maximum value if
$\varphi_P-\varphi_{S_1}-\varphi_{I_1}=\mbox{arg}(\xi_P\xi_{S_1}^\ast
\xi_{I_1}^\ast)=-\frac{\pi}{2}$ (see Fig~\ref{fig45}~(b)). It is also
evident from the second term of the R.H.S.
of Eq.~(\ref{shortcorelS2I2}) that the linear interaction enables us to affect
the anti-correlation in mode $(S_2,I_2)$ via amplitudes $\xi_{S_1}$,
$\xi_{I_1}$.

Repeating the arguments leading to the formula (\ref{corelS2I2}) for
mode $(S_1,I_1)$, we obtain
%%%%%%%%%%%%%%%%%%%%%%%%%%%%%%%%%%%%%%%%%%%%%%%%%%%%%%%%%%%%%%%%%%%%%
\begin{eqnarray}\label{corelS1I1}
\langle\Delta W_{S_1}(z)\Delta W_{I_1}(z)\rangle&=&|G|^2
\left[2\left(|\xi_{S_1}|^2+|\xi_{I_1}|^2\right)+1\right]u^2(z)v^2(z)
\nonumber\\
&&-2|G||\xi_{S_1}||\xi_{I_1}|\,\mbox{sin}(\varphi_P-\varphi_{S_1}-
\varphi_{I_1})\nonumber\\
&&\times u(z)v(z)\left[u^2(z)+|G|^2v^2(z)\right].
\end{eqnarray}
%%%%%%%%%%%%%%%%%%%%%%%%%%%%%%%%%%%%%%%%%%%%%%%%%%%%%%%%%%%%%%%%%%%%%
Employing once more the expansion of (\ref{uvfce}) around $z=0$, we can see,
that the second term on the R.H.S. of Eq.~(\ref{corelS1I1}) is proportional
to $z$ and thus its sign is given only by the argument of sine function.
Our choice of the initial phases then implies that the contribution of the
second term is positive in this approximation. However, the product $u(z)v(z)$
alternates and its amplitude increases with increasing $z$, suppressing the
quantum noise represented by the first term on the R.H.S. of
Eq.~(\ref{corelS1I1}) and attaining the sub-Poissonian photon
statistics for larger $z$.
%%%%%%%%%%%%%%%%%%%%figure6%%%%%%%%%%%%%%%%%%%%%%%%%%%%%%%%%
%%%%%%%%%%%%%%%%%%%%%%%%%%%%%%%%%%%%%%%%%%%%%%%%%%%%%%%%%%%%%%%%%%%
\subsection{\it Cross mode}

Up to now we have separately discussed modes localized either in the
first or in the second waveguide. This subsection will be devoted to
the investigation of non-classical behaviour occuring in cross mode
$(S_1,I_2)$.
%%%%%%%%%%%%%%%%%%%%%%%%figures7(a),(b)%%%%%%%%%%%%%%%%%%%%%%%%%%%%%
\begin{figure}
\vspace{-4cm}
\centerline{\hspace{0cm}\psfig{width=17cm,angle=0,file=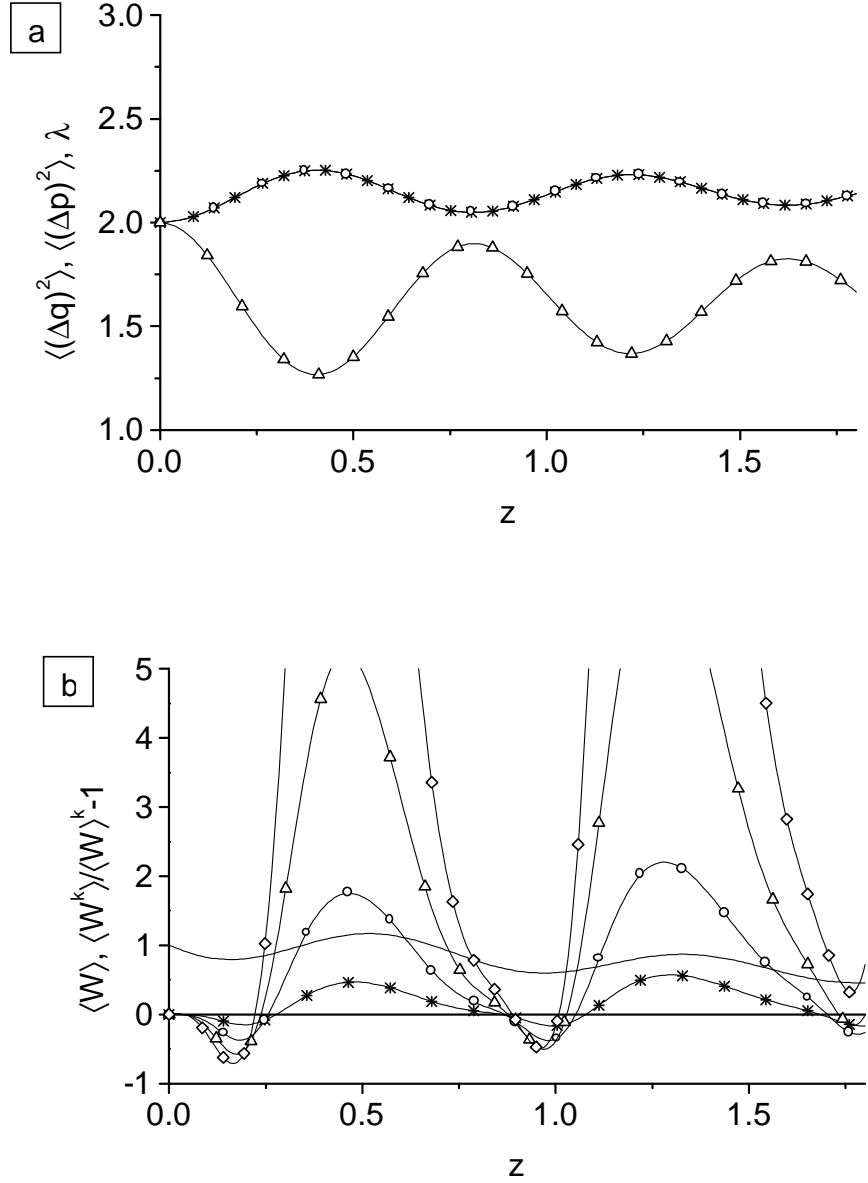}}
\vspace{-5cm}
\caption{(a) The quadrature variances $\langle{\left[\Delta\hat{q}(z)
\right]}^2\rangle$ $(\circ)$, $\langle{\left[\Delta\hat{p}(z)\right]}^2
\rangle$ $(\ast)$ and principal squeeze variance $\lambda(z)$ $(\triangle)$
for mode $(S_1,I_2)$, (b) integrated intensity $\langle W(z)\rangle$ (---)
and its reduced factorial moments $\frac{\langle W^k(z)\rangle}
{{\langle W(z)\rangle}^k}-1$ for $k=2$ $(\ast)$, $k=3$ $(\circ)$, $k=4$
$(\triangle)$, $k=5$ $(\diamond)$ for mode $(S_1,I_2)$; $\Gamma_1=\Gamma_2=1$,
$\kappa_S=\kappa_I=4$, $\Delta k=\Delta k_{S}=\Delta k_{I}=0$, $\xi_{P_1}=
\xi_{P_2}=i$, $\xi_{S_1}=\xi_{I_1}=1$, $\xi_{S_2}=\xi_{I_2}=0$,
$\gamma_j=0.2$, $\langle n_{dj}\rangle=10^{-2}$.}
\label{fig78}
\end{figure}
%%%%%%%%%%%%%%%%%%%%%%%%%%%%%%%%%%%%%%%%%%%%%%%%%%%%%%%%%%%%%%%%%%%%%%%%
The investigation of the down-conversion with strong pumping led to the
conclusion that signal mode $S$ and idler mode $I$ do not exhibit any
non-classical behaviour irrespectively of the fact, if they are spontaneous
or stimulated by the coherent light \cite{Permon}. Obviously, mode
$(S_1,I_2)$ compounded of modes $S_1$ and $I_2$ originating from two
independent down-conversion processes cannot provide a non-classical
light either. However, introducing the linear interaction between the processes,
we can observe squeezing of vacuum fluctuations and sub-Poissonian photon
statistics simultaneously in this mode (see Figure~\ref{fig78}).
To discover the origin of these phenomena we will employ
once more the analytical solution. In the spirit of the derivation of
Eq.~(\ref{corelS2I2}) we can calculate the cross-correlation function
%%%%%%%%%%%%%%%%%%%%%%%%%%%%%%%%%%%%%%%%%%%%%%%%%%%%%%%%%%%%%%%%%%%%
\begin{equation}\label{corelf}
D_{S_1I_2}(z)=-\kappa|G|v^2(z),
\end{equation}
%%%%%%%%%%%%%%%%%%%%%%%%%%%%%%%%%%%%%%%%%%%%%%%%%%%%%%%%%%%%%%%%%%%%
indicating, that linear exchange introduces the correlation between modes
$S_1$ and $I_2$. This correlation reduces both vacuum fluctuations
(see (\ref{lambda})) and fluctuations of photon number. To make the
latter more clear, we can derive the following cross-correlation function
up to $z^3$
%%%%%%%%%%%%%%%%%%%%%%%%%%%%%%%%%%%%%%%%%%%%%%%%%%%%%%%%%%%%%%%%%%%%
\begin{equation}\label{shortcorelS1I2}
\langle\Delta W_{S_1}(z)\Delta W_{I_2}(z)\rangle\approx-2\kappa^2|G|
|\xi_{S_1}||\xi_{I_1}|z^3\mbox{sin}(\varphi_P-\varphi_{S_1}-
\varphi_{I_1}).
\end{equation}
%%%%%%%%%%%%%%%%%%%%%%%%%%%%%%%%%%%%%%%%%%%%%%%%%%%%%%%%%%%%%%%%%%%%
It is worth noting that the effect of noise reduction can be enhanced by
increasing the amplitudes $\xi_{S_1}$ and $\xi_{I_1}$.

%%%%%%%%%%%%%%%%%%%%%%%%%%%%%%%%%%%%%%%%%%%%%%%%%%%%%%%%%%%%%%%%%%%%
\subsection{\it Mismatch-controlled switching}

\begin{figure}
\vspace{-6cm}
\centerline{\hspace{0cm}\psfig{width=17cm,angle=0,file=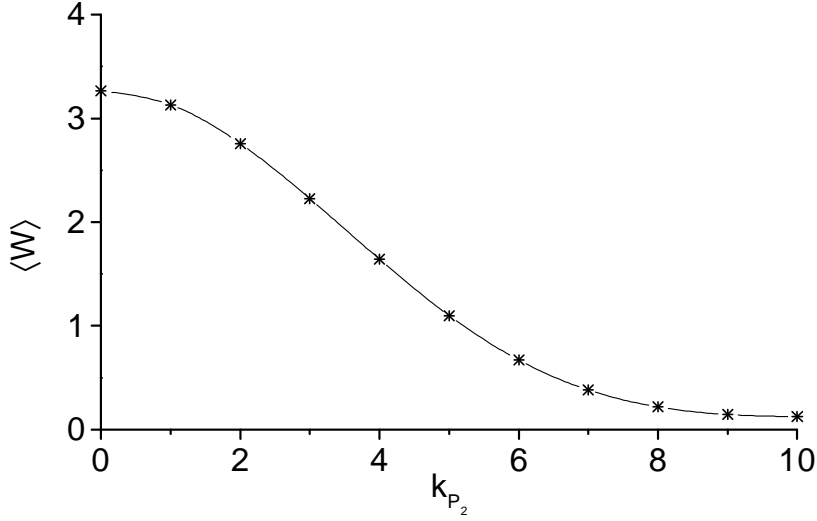}}
\vspace{-10.5cm}
\caption{The dependence of the integrated intensity $\langle W(z)\rangle$
of $(S_1,I_1)$ (---) and $(S_2,I_2)$ $(\ast)$ modes on the wavevector
along the $z$-axis of propagation $k_{P_2}$ of pump mode $P_2$; $L=1.5$,
$\Gamma_1=\Gamma_2=1$, $\kappa_S=\kappa_I=2.5$, $k_{S_1}=k_{S_2}=6$,
$k_{I_1}=k_{I_2}=4$, $k_{P_1}=10$, $\xi_{P_1}=\xi_{P_2}=1$, $\xi_{S_1}=
\xi_{I_1}=\xi_{S_2}=\xi_{I_2}=0$, $\gamma_j=0.2$, $\langle n_{dj}\rangle=
10^{-2}$.}
\label{fig_zeno}
\end{figure}

Before discussing the last phenomenon we would like to mention several
general remarks concerning the all-optical switching. This will enlight
the motivation of the following discussion. Recent theoretical
investigation of couplers has led to an interesting conclusion. Not only can
they serve as a passive optical switchers, but they also provide
the active control of the output of a particular waveguide by means of the
input of the other one. There are at least two ways how to actively control
the output beams. First, the coupling lenght of the coupler can be
adjusted by changing the intensity of the strong classical input
field \cite{Asant,Janszky}. Second, the phase-controlled distribution of
the quantum noise in couplers can be realized \cite{Mista2} (see also
Subection~\ref{sub_swit}). There is, however, one more possibility how to
affect the properties of the outgoing beams. Inspection of Fig.~\ref{fig_V}
reveals that one can change the dynamical behaviour of the beams by means of
the global mismatch $\Delta k$. Moreover, due to its global character
(it contains all wavevectors), we can control one mode by means of another one,
even though they directly do not interact. The following arrangement
can illustrate this. Let us assume that both processes are spontaneous,
nonzero wavevectors $k_{S_1}$, $k_{S_2}$, $k_{I_1}$, $k_{I_2}$ and $k_{P_1}$
are chosen in such a way, that they satisfy the matching conditions
$(\Delta k_S=\Delta k_I=\Delta l_1=0)$ and linear interaction is in operation.
Now the increase of the $z$-th component $k_{P_2}$ of the wavevector of
mode $P_2$ entails the inhibition of the decay of the pump photons in the
first waveguide (see Figure~\ref{fig_zeno}). This phenomenon is easy to
explain based on Fig.~\ref{fig_V}. Initially, the parameters of the
coupler are such that the down-convertion part of the evolution is dominant
(we are inside the hatched area). The global mismatch $\Delta k$ decreases
with increasing $k_{P_2}$ and the linear part of evolution grows dominant.
Interpreting once again the linear interaction as a sort of continuous
measurement, this effect can be looked at as a complementary effect
to the Zeno or anti-Zeno-like effects described in Sec.~\ref{sec_dyn}.
Unlike in Sec.~\ref{sec_dyn} where initial condition (the value of
$\Delta k$) was kept constant and the strength of ``measurement''
was changed, here the strength of the ``measurement'' is the fixed
quantity and the initial condition is continuously varied. In this way
the influence of the ``measurement'' results in the speeding up the
down-conversion for small values $k_{P_2}$ and slowing down the
down-conversion for larger $k_{P_2}$. This corresponds to a transition
from the anti-Zeno to Zeno regime.
%%%%%%%%%%%%%%%%%%%%%%%%figure9%%%%%%%%%%%%%%%%%%%%%%%%%%%%%%%%%%%%
%%%%%%%%%%%%%%%%%%%%%%%%%%%%%%%%%%%%%%%%%%%%%%%%%%%%%%%%%%%%%%%%%%%%%%

It is also worth noting, that the integrated intensity of mode $(S_2,I_2)$
depends on $k_{P_2}$ in the same way as mode $(S_1,I_1)$ does (see
Fig.~\ref{fig_zeno}). This can be understood as follows. At the beginning
(when $k_{P_2}=0$) the process in the first waveguide is perfectly matched
$(\Delta l_1=0)$ and the process in the second waveguide is strongly
mismatched $(\Delta l_2\not=0)$. The linear interaction, however, symmetrizes
the device in the way that it partially mismatches the first process and
partially compensates the mismatch in the second waveguide at the same
time.
%%%%%%%%%%%%%%%%%%%%%%%%%%%%%%%%%%%%%%%%%%%%%%%%%%%%%%%%%%%%%%%%%%%%
\section{Conclusion}\label{sec_con}

The quantum dynamics and statistics of the symmetric nonlinear coupler
operating by down-conversion process have been investigated. In a
framework of strong pumping approximation we have solved analytically the
Heisenberg-Langevin equations. The manifestation of Zeno and anti-Zeno
effects has been demonstrated based on the analytical solution. The
non-classical behaviour of beams involved has been studied based on
numerical calculations. The phase-controlled redistribution of quantum
noise between the quadratures can be achieved in mode $(S_1,I_1)$.
The possibility of generation of sub-Poissonian light in modes $(S_1,I_1)$
and $(S_2,I_2)$ caused by the linear interaction of two super-Poissonian
lights has been shown. Light exhibiting simultaneous squeezing of
vacuum fluctuations and sub-Poissonian photon statistics can be
obtained in cross mode $(S_1,I_2)$. The inhibition of the decay process
in the first waveguide owing to the nonlinear matching of the second
process has been observed. All these phenomena were shown to be robust
against the presence of weak damping.

\appendix
\section{Matrices $\bf b$, $\bf c$, $\bf d$ of Eq.~(\ref{amatrix})}
\label{appA}
%%%%%%%%%%%%%%%%%%%%%%%%%%%%%%%%%%%%%%%%%%%%%%%%%%%%%%%%%%%%%%%%%%%%
\small
\begin{displaymath}\label{bmatrix}
{\bf b}=
\left[\begin{array}{cccc}
K_{I_1}+K_{S_2}+K_{I_2} & i\kappa_S^\ast & iG_1 & 0  \\
i\kappa_S & K_{S_1}+K_{I_1}+K_{I_2} & 0 & iG_2  \\
-iG_1^\ast & 0 & K_{S_1}+K_{S_2}+K_{I_2} & -i\kappa_I  \\
 0 & -iG_2^\ast & -i\kappa_I^\ast & K_{S_1}+K_{S_2}+K_{I_1}
\end{array} \right],\nonumber\\
\end{displaymath}
%%%%%%%%%%%%%%%%%%%%%%%%%%%%%%%%%%%%%%%%%%%%%%%
\small
\begin{eqnarray*}\label{cmatrix}
{\bf c}=
\left[\begin{array}{ccc}
K_{I_1}K_{S_2}+L_I+\bar L_2 & i{\kappa_S^\ast}(K_{I_1}+K_{I_2}) &
iG_1(K_{S_2}+K_{I_2}) \\ 
i{\kappa_S}(K_{I_1}+K_{I_2}) & K_{S_1}K_{I_2}+L_I+\bar L_1 &
{\kappa_I^\ast} G_2-{\kappa_S}G_1 \\ 
-iG_1^\ast(K_{S_2}+K_{I_2}) & {\kappa_S^\ast} G_1^\ast
-{\kappa_I}G_2^\ast & K_{S_1}K_{I_2}+L_S+\bar L_2 \\
{\kappa_S}G_2^\ast-{\kappa_I^\ast} G_1^\ast &
-iG_2^\ast(K_{S_1}+K_{I_1}) & -i\kappa_I^\ast(K_{S_1}+K_{S_2}) \\
\end{array} \right.\\
\left.\begin{array}{c}
{\kappa_I}G_1-{\kappa_S^\ast} G_2 \\ iG_2(K_{S_1}+K_{I_1})  \\
-i\kappa_I(K_{S_1}+K_{S_2}) \\ K_{I_1}K_{S_2}+L_S+\bar L_1 \\
\end{array} \right],
\end{eqnarray*}
%%%%%%%%%%%%%%%%%%%%%%%%%%%%%%%%%%%%%%%%%%%%%%%
\begin{eqnarray*}\label{dmatrix}
{\bf d}=
\left[\begin{array}{ccc}
{\bar L_2}K_{I_1}+|\kappa_I|^2K_{S_2} & i\kappa_S^\ast L_I
-i\kappa_I G_1G_2^\ast & iG_1{\bar L_2}+i{\kappa_S^\ast}{\kappa_I^\ast}
G_2 \\
i{\kappa_S}L_I-i\kappa_I^\ast G_1^\ast G_2 & {\bar L_1}K_{I_2}+
|\kappa_I|^2K_{S_1} & {\kappa_I^\ast} G_2K_{S_1}-{\kappa_S}G_1K_{I_2} \\
-iG_1^\ast{\bar L_2}-i\kappa_S\kappa_IG_2^\ast & {\kappa_S^\ast} G_1^\ast
K_{I_2}-{\kappa_I}G_2^\ast K_{S_1} & {\bar
L_2}K_{S_1}+|\kappa_S|^2K_{I_2} \\
\kappa_S G_2^\ast K_{I_1}-{\kappa_I^\ast} G_1^\ast K_{S_2} &
-i{G_2^\ast}{\bar L_1}-i{\kappa_S^\ast}{\kappa_I^\ast} G_1^\ast
& -i\kappa_I^\ast L_S+i\kappa_S G_1 G_2^\ast \\
\end{array} \right. \\
\left.\begin{array}{c}
-{\kappa_S^\ast} G_2 K_{I_1}+{\kappa_I}G_1 K_{S_2} \\
iG_2{\bar L_1}+i{\kappa_S}{\kappa_I}G_1  \\
-i\kappa_I L_S+i{\kappa_S^\ast}G_1^\ast G_2 \\
{\bar L_1}K_{S_2}+|\kappa_S|^2K_{I_1}
\end{array} \right].\nonumber\\
\end{eqnarray*}
\normalsize
%%%%%%%%%%%%%%%%%%%%%%%%%%%%%%%%%%%%%%%%%%%%%%%%%%%%%%%%%%%%%%
\section{Noise functions}
\label{appB}
%%%%%%%%%%%%%%%%%%%%%%%%%%%%%%%%%%%%%%%%%%%%%%%%%%%%%%%%%%%%%%
\begin{eqnarray}\label{noise}
B_{S_1}(z)&=&\langle\Delta\hat{A}_{S_1}^\dag(z)\Delta\hat{A}_{S_1}
(z)\rangle=\sum_{j=1}^{4}\left(|X_{1j}|^2 B_j+
2\gamma_j\langle{n_{dj}}\rangle\chi_{1j}\right)\nonumber\\
&&+\sum_{j=1}^{2}\left(2\gamma_{j+2}\chi_{1j+2}-|X_{1j}|^2\right),
\nonumber\\
C_{S_1}(z)&=&\langle(\Delta\hat{A}_{S_1}(z))^2\rangle=\sum_{j=1}^{2}\left(
X_{1j}^2C_j+X_{1j+2}^2C_{j+2}^\ast\right)\mbox{exp}(-2i\Delta K_{S_1}z),
\nonumber\\
D_{S_1S_2}(z)&=&\langle\Delta\hat{A}_{S_1}(z)\Delta\hat{A}_{S_2}
(z)\rangle=\sum_{j=1}^{2}\left(X_{1j}X_{2j}C_j+X_{1j+2}X_{2j+2}C_{j+2}^\ast
\right)\nonumber\\
&&\times\mbox{exp}(-i\Delta k),
\nonumber\\
D_{S_1I_1}(z)&=&\langle\Delta\hat{A}_{S_1}(z)\Delta\hat{A}_{I_1}(z)\rangle=
\left[\sum_{j=1}^{4}\left(X_{1j}X_{3j}^\ast B_j+2\gamma_j\langle{n_{dj}}
\rangle\chi_{jj}^{13}\right)\right.\nonumber\\
&&\left.+\sum_{j=1}^{2}\left(2\gamma_j\chi_{jj}^{13}-X_{1j+2}X_{3j+2}^\ast
\right)\right]\mbox{exp}[-i(\Delta K_{S_1}+\Delta K_{I_1})z],
\nonumber\\
D_{S_1I_2}(z)&=&\langle\Delta\hat{A}_{S_1}(z)\Delta\hat{A}_{I_2}(z)\rangle=
\left[\sum_{j=1}^{4}\left(X_{1j}X_{4j}^\ast B_j+2\gamma_j\langle{n_{dj}}
\rangle\chi_{jj}^{14}\right)\right.\nonumber\\
&&\left.+\sum_{j=1}^{2}\left(2\gamma_j\chi_{jj}^{14}-X_{1j+2}X_{4j+2}^\ast
\right)\right]\mbox{exp}[-i(\Delta K_{S_1}+\Delta K_{I_2})z],\nonumber\\
\bar{D}_{S_1S_2}(z)&=&-\langle\Delta\hat{A}_{S_1}^\dag(z)\Delta\hat{A}_{S_2}(z)
\rangle=-\left[\sum_{j=1}^{4}\left(X_{1j}^\ast X_{2j}B_j
+2\gamma_j\langle{n_{dj}}\rangle\chi_{jj}^{21}\right)\right.\nonumber\\
&&\left.+\sum_{j=1}^{2}\left(2\gamma_{j+2}\chi_{j+2j+2}^{21}-X_{1j}^\ast
X_{2j}\right)\right]\mbox{exp}(-i\Delta k_S z),\nonumber\\
\bar{D}_{S_1I_1}(z)&=&-\langle\Delta\hat{A}_{S_1}^\dag(z)\Delta\hat{A}_{I_1}
(z)\rangle=-\sum_{j=1}^{2}\left(X_{1j}^\ast X_{3j}^\ast C_j^\ast+X_{1j+2}^\ast
X_{3j+2}^\ast C_{j+2}\right)\nonumber\\
&&\times\mbox{exp}[i(\Delta K_{S_1}-\Delta K_{I_1})z],
\nonumber\\
\bar{D}_{S_1I_2}(z)&=&-\langle\Delta\hat{A}_{S_1}^\dag(z)\Delta\hat{A}_{I_2}(z)
\rangle=-\sum_{j=1}^{2}\left(X_{1j}^\ast X_{4j}^\ast C_j^\ast+X_{1j+2}^\ast
X_{4j+2}^\ast C_{j+2}\right)\nonumber\\
&&\times\mbox{exp}[i(\Delta K_{S_1}-\Delta K_{I_2})z],
\end{eqnarray}
%%%%%%%%%%%%%%%%%%%%%%%%%%%%%%%%%%%%%%%%%%%%%%%%%%%%%%%%%%%
where $X_{ij}=X_{ij}(z)$ are defined in Eq.~(\ref{xmatrix}) and
%%%%%%%%%%%%%%%%%%%%%%%%%%%%%%%%%%%%%%%%%%%%%%%%%%%%%%%%%%%
\begin{eqnarray*}\label{chi}
\chi_{ij}&=&\chi_{ij}(z)=\int_{0}^{z}|X_{ij}(z-z')|^2dz',\\
\chi_{jl}^{ik}&=&\chi_{jl}^{ik}(z)=\int_{0}^{z}X_{ij}(z-z')
X_{kl}^\ast(z-z')dz'.
\end{eqnarray*}
%%%%%%%%%%%%%%%%%%%%%%%%%%%%%%%%%%%%%%%%%%%%%%%%%%%%%%%%%%%
The rest of the noise functions can be obtained using the symmetry of
the model.
%%%%%%%%%%%%%%%%%%%%%%%%%%%%%%%%%%%%%%%%%%%%%%%%%%%%%%%%%%%
\\[12pt]
\noindent
{\large \bf {Acknowledgments}}\\
We would like to thank J. Pe\v{r}ina Jr. for help with numerical calculations.
Support by Grant No. VS96028, Research project CEZ:J14 "Wave and Particle
Optics" of the Czech Ministry of Education and Grant No. 202/00/0142 of
Czech Grant Agency is acknowledged.
%%%%%%%%%%%%%%%%%%%%%%%%%%%%%%%%%%%%%%%%%%%%%%%%%%%%%%%%%%%

\end{document}